\documentclass[11pt]{article} 
\usepackage[margin=1.0in]{geometry}
\usepackage{amsmath,amssymb}
\usepackage{graphicx}
\graphicspath{{figures/}}
\usepackage{color}
\usepackage{lineno}
\usepackage{color} 
\usepackage{setspace}

\newcommand{\ra}{\rightarrow}

\newcommand{\lra}{\leftrightarrows}

\newcommand{\vtx}[1]{\langle\, #1 \,\rangle}
\newcommand{\lap}{\mbox{$\cal L$}}

\newcommand{\pr}{\mbox{Pr}}
\newcommand{\ol}[1]{\overline{#1}}
\newcommand{\eps}{\varepsilon}
\newcommand{\email}[1]{\tt{#1}}

\spacing{1}
\makeatletter
\renewcommand{\maketitle}{\bgroup\setlength{\parindent}{0pt}
\begin{flushleft}
\textbf{\@title}
\@author
\end{flushleft}\egroup
}
\makeatother
\title{\Large Reformulating non-equilibrium steady-states and generalised Hopfield discrimination}
\date{}
\author{\vspace{8pt}\\
U\u{g}ur \c{C}etiner$^{1}$, 
and 
Jeremy Gunawardena$^{1,\dagger}$\\
}

\begin{document}
\maketitle

{\small
$^{1}$Department of Systems Biology, Harvard Medical School, Boston, MA 02111, USA

$^{\dagger}$Corresponding author: Jeremy Gunawardena \email{(jeremy@hms.harvard.edu)}
}

\vspace{0.6in}
{\noindent
ABSTRACT \\[1em]
\textbf{Despite substantial progress in non-equilibrium physics, steady-state (s.s.) probabilities remain intractable to analysis. For a Markov process, s.s. probabilities can be expressed in terms of transition rates using the Matrix-Tree theorem (MTT) in the graph-based linear framework. The MTT reveals that, away from equilibrium, s.s. probabilities become globally dependent on all rates, with expressions growing exponentially in the system size. This overwhelming complexity and lack of thermodynamic interpretation have greatly impeded analysis. Here, we show that s.s. probabilities are proportional to the average of $\exp(-S(P))$, where $S(P)$ is the entropy generated along minimal paths, $P$, in the graph, and the average is taken over a probability distribution on spanning trees. Assuming Arrhenius rates, this ``arboreal'' distribution becomes Boltzmann-like, with  the energy of a tree being its total edge barrier energy. This reformulation offers a thermodynamic interpretation that smoothly generalises equilibrium statistical mechanics and reorganises the expression complexity: the number of distinct minimal-path entropies depends on the entropy production index, a new graph-theoretic measure of non-equilibrium complexity, not on graph size. We demonstrate the power of this reformulation by extending Hopfield's analysis of discrimination by kinetic proofreading to any graph with index 1. We derive a general formula for the error ratio and use it to show that local energy dissipation can yield optimal discrimination through global synergy.}
\section*{INTRODUCTION}
Equilibrium thermodynamics and statistical mechanics are among the great successes of 19th century physics and remain essential for studies in many fields. In contrast, despite impressive advances, the foundations of non-equilibrium physics remain under development. This gap has had significant repercussions in biology, since life itself is quintessentially far from equilibrium. Although much is known about the molecular components involved in energy transduction, the functional significance of energy expenditure has been harder to unravel, especially for cellular information processing. 

The biophysicist Terrell Hill introduced in the 1960s an approach to analysing individual non-equilibrium entities, such as a membrane transporter or a motor protein, based on mesocopic states and transitions represented in ``diagrammatic'' form \cite{hill66,hill04}. In essence, this was a Markov process described by a graph. J\u{u}rgen Schnakenberg developed this approach further in the 1970s in his ``network theory'' \cite{sch76}. An important contribution of these studies was to show how graph cycles related macroscopic thermodynamic quantities like entropy production to the underlying stochastic mesoscopic quantities. For reasons that remain unclear, this graph-based approach then disappeared from sight in the physics literature. In particular, it played no role in the renaissance of non-equilibrium statistical mechanics which began in the 1990s and has led to exact fluctuation theorems for systems arbitrarily far from thermodynamic equilibrium \cite{ecm93,gec95,jar97b,kur98,lhs99,mae99,cro99}. As physicists began to build on these new findings, Markov processes became a foundational tool for stochastic thermodynamics \cite{sei08} and their graph-based representations began to be rediscovered \cite{aga07,mkn13,mhl14}.

Graphs also make an appearance in the pioneering work of Wentzell and Freidlin in large deviation theory \cite{vmf70,frewen}. Here, the graph offers a discrete approximation to a stochastic differential equation in the limit of low noise or low temperature. Vertices correspond to stable steady states, edges to appropriate barrier crossings and labels to crossing rates. Such graph-theoretic approximations have been further developed within chemical physics, especially for analysing complex free-energy landscapes at thermodynamic equilibrium \cite{djw06,cve14}. 

Independently of these developments, a graph-theoretic approach to analysing biochemical systems under timescale separation, the ``linear framework'', was introduced in systems biology \cite{gun-mt,jg-inom-lapd,mbo15,yst18,yst20}. This was applied both to bulk populations of biochemical entities, such as post-translational modification systems \cite{dcg12,nbg20}, and individual stochastic entities, such as a gene \cite{edg16,bng18,wcg18,pgd18,wjg19}. In the latter context, as in the approaches described above, the linear framework provides a treatment of continuous-time, finite state Markov processes based on directed graphs with labelled edges. Vertices correspond to mesostates, directed edges to transitions and edge labels to transition rates. The main distinction with the approaches described above is that the graph is treated as a mathematical entity in its own right. This offers a rigorous way to relate network structure to function that is well-suited to rising above the molecular complexity found in biology \cite{wag18,wcg18}. 

The graph-theoretic approach offers particular insight into a problem which has resisted the breakthroughs in non-equilibrium statistical mechanics mentioned above: the probabilities of mesoscopic states, even at steady-state, remain intractable to exact analysis. We explain the issues here in broad terms before giving full details below. 

The s.s. probabilities of a Markov process can be expressed in terms of its graph edge labels by using certain subgraphs---spanning trees---as described by the Matrix-Tree theorem (MTT, Eq.\ref{e-mtt}). Results of this kind date back to Kirchhoff \cite{gki47}. The version used here was first stated by Tutte \cite{tut48} but independently rediscovered by Hill \cite{hill66} and by many others \cite{jg-inom-lapd}. The MTT makes clear that, as soon as the system is away from thermodynamic equilibrium, even if that occurs through energy expenditure at only a single edge, the s.s. probabilities become globally dependent on all edge labels in the graph. The resulting expressions, which depend on enumerating all spanning trees, become extremely complex. Consider, for example, a graph whose mesostates correspond to the presence or absence at $k$ sites of some feature, such as a post-translational modification, so that there are $2^k$ mesostates. For $k = 2$, there are 4 spanning trees; for $k = 3$ there are 384 spanning trees; but for $k = 4$ there are 42,467,328 spanning trees \cite{edg16}. All of these trees are required to exactly determine s.s. probabilities. Moreover, while the mathematical details are clear, a thermodynamic interpretation of these expressions has been lacking. We have not been able to see the wood for the trees. The combinatorial complexity and lack of thermodynamic meaning have greatly hindered exact calculations, even for systems which are, from an equilibrium perspective, very straightforward. 

We offer a solution to both these challenges. Since the complexity cannot be avoided, it must be reorganised and reinterpreted. There are two parts to this reformulation. First, we focus on minimal paths in the graph from a given vertex, $i$ to a reference vertex, $1$. Minimal paths are those with no repeated vertices. There are only finitely many minimal paths in a finite graph. Let $S(P)$ denote the overall entropy production from taking the path $P$. Detailed balance tells us that the equilibrium s.s. probability of vertex $i$ is proportional to $\exp(-S(P))$, no matter which path $P$ is chosen from $i$ to $1$. This is equilibrium statistical mechanics in the graph setting. Second, to address the situation away from equilibrium, we define a probability distribution on spanning trees rooted at $1$. We call this the ``arboreal distribution''. To provide a thermodynamic interpretation, we write edge labels in Arrhenius form, in terms of a vertex energy and an edge barrier energy; our constructions are independent of the choices involved. Surprisingly, in view of the non-equilibrium setting, the arboreal distribution is Boltzmann-like, with $\pr(T) \propto e^{-E(T)}$, where the energy $E(T)$ is the sum of the barrier energies on all edges of the tree $T$ (Eq.\ref{e-et}). Each spanning tree rooted at $1$ yields a unique path from $i$ to $1$, for each vertex $i$. Our main result is that, away from thermodynamic equilibrium, the s.s. probability of vertex $i$ is proportional to the average of $\exp(-S(P))$, where $P$ runs over the minimal paths from $i$ to $1$ and the average is taken over the arboreal distribution (Eq.\ref{e-mt3}). 

This reformulation provides a thermodynamic interpretation in place of a forest of trees and smoothly generalises equilibrium statistical mechanics. It also finesses the combinatorial explosion: the number of distinct values for the entropy production on minimal paths has a different scaling to the number of spanning trees or the number of minimal paths themselves. The scaling does not depend on the size of the graph; rather it depends on how many edges in the graph are experiencing energy expenditure (below). This revised scaling dramatically simplifies the calculation of s.s. probabilities. 

To illustrate the power of this reformulation we substantially generalise Hopfield's classic study of discrimination by kinetic proofreading \cite{hop74}. Hopfield analysed a simple graph with 3 vertices. We analyse any graph in which energy is expended at only one edge and give a general formula for the error ratio (Eq.\ref{e-epq}). We exploit this formula to show that optimal discrimination is possible even in complex graphs, despite energy being expended at only one edge.

Finally, we introduce the entropy production index as a measure of departure from thermodynamic equilibrium. One of the messages of this paper is that systems whose index is one, which corresponds to energy expenditure at a single edge, although being away from equilibrium and suffering all the problems of global parameter dependence and combinatorial complexity are nevertheless algebraically tractable through the reformulation of steady-state probabilities presented here. 

\section*{RESULTS}

\subsection*{Steady-state probabilities in the linear framework}

We briefly describe the linear framework. More details and background can be found in \cite{gun-mt,jg-inom-lapd,edg16,bng18,wjg19}. Let $G$ denote a finite, directed graph with labelled edges and no self-loops (Fig.\ref{f-1}A). We denote the vertices of $G$ by the indices, $1, 2, \cdots, n$, an edge from $i$ to $j$ by $i \ra j$ and the label on this edge by $\ell(i \ra j)$. We think of the vertices as mesostates of the system under study, implying thereby that they are coarse-grained abstractions of the underlying physical microstates. The edges correspond to transitions between mesostates with the label being the transition rate, with dimensions of (time)$^{-1}$. Labels may be complex expressions which describe interactions between mesostates and environmental reservoirs, such as those for molecular entities (particles) or heat. We make the customary thermodynamic assumption that exchanges between the graph and the reservoirs, for instance through binding or unbinding of a ligand, do not change the thermodynamic potentials of the reservoirs. Edge labels may then be treated as constants.

\begin{figure}[t]
\begin{center}
\includegraphics[viewport=56 538 508 746,clip=true,width=0.75\textwidth,keepaspectratio]{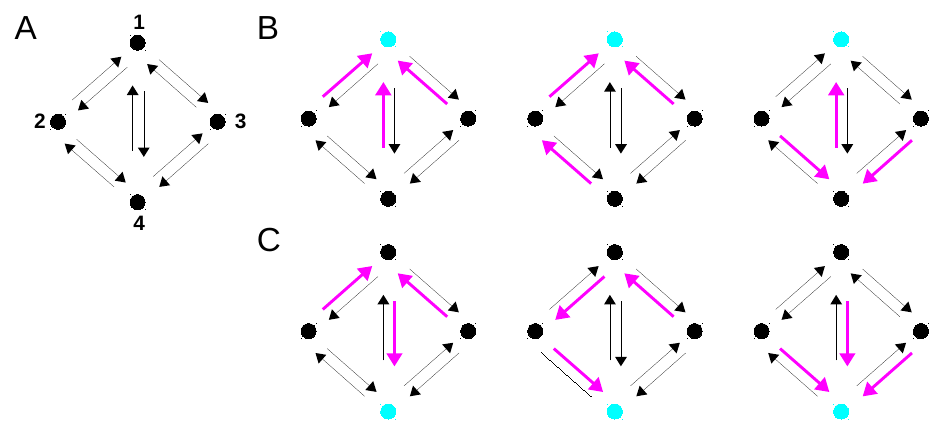}  
\caption{Graphs and spanning trees. \textbf{A} Reversible linear framework graph, with named vertices and labels omitted. \textbf{B} Three spanning trees (magenta edges) rooted at vertex $1$ (cyan), chosen from 8 possibilities. \textbf{C} Corresponding spanning trees rooted at vertex $4$, obtained by applying the map $\Phi_{1,4}$, as described in the text, to the tree vertically above in panel \textbf{B}, with the same colour code as \textbf{B}.}
\label{f-1}
\end{center}
\end{figure} 

$G$ describes the infinitesimal generator of a continuous-time Markov process, $X(t)$, given by a conditional probability distribution on the same mesostates for times $s > t$, $\pr(X(s) = j \,|\, X(t) = i)$. The edge labels are those infinitesimal transition rates, 
\begin{equation}
\ell(i \ra j) = \lim_{\Delta t \ra 0} \frac{\pr(X(t + \Delta t) = j \,|\, X(t) = i)}{\Delta t} \,,
\label{e-mkp}
\end{equation}
which are not zero. Provided the limits in Eq.\ref{e-mkp} exist, there is an exact correspondence between Markov processes and graph representations \cite{jg-inom-lapd}. In particular, the master equation of the Markov process, which describes the deterministic time evolution of the probabilities of mesostates, can be recovered from the graph. Let $u_i(t)$ denote the probability of mesostate $i$ at time $t$, $u_i(t) = \pr(X(t) = i)$. The master equation is the linear matrix equation
\begin{equation}
\frac{du(t)}{dt} = \lap(G)u(t) \,,
\label{e-lap}
\end{equation}
where $\lap(G)$ is the $n \times n$ Laplacian matrix of $G$ \cite{gun-mt}.

Since Eq.\ref{e-lap} is linear, there is no difficulty in solving it in terms of eigenvalues but these are not known in terms of the edge labels, at least for $n \geq 5$. However, the s.s. probabilities of the mesostates, denoted $u^*(G)$, can be expressed in terms of the labels. If $H$ is a subgraph of $G$, let $q(H)$ denote the product of the labels on the edges of $H$: $q(H) = \prod_{i \ra j \in H} \ell(i \ra j)$. Let $\Theta_i(G)$ denote the set of spanning trees of $G$ rooted at $i$. A spanning tree is a subgraph which includes each vertex of $G$ (spanning) and has no cycles if edge directions are ignored (tree); it is rooted at $i$ if the tree has no edges outgoing from $i$ (Fig.\ref{f-1}B). Provided $G$ is strongly connected, so that any two vertices, $i$ and $j$, are joined by a directed path, $i = i_1 \ra i_2 \ra \cdots \ra i_k = j$, there exist rooted spanning trees at each vertex. Moreover, the kernel of $\lap(G)$ is one dimensional. A canonical basis element, $\rho(G) \in \ker\lap(G)$, is given by the Matrix-Tree theorem (MTT),
\begin{equation}
\rho_i(G) = \sum_{T \in \Theta_i(G)} q(T) \,.
\label{e-mtt}
\end{equation}
Since $u^*(G) \in \ker\lap(G)$, it follows that $u^*(G) \propto \rho(G)$. The proportionality constant comes from solving for total probability, $u^*_1(G) + \cdots + u^*_n(G)= 1$, which gives,
\begin{equation}
u^*_i(G) = \frac{\rho_i(G)}{\rho_1(G) + \cdots + \rho_n(G)} \,.
\label{e-ssp}
\end{equation}

\subsection*{Path entropies and thermodynamic equilibrium}

We assume from now on that $G$ is reversible: if $i \ra j$, then also $j \ra i$, and, furthermore, the reverse edge represents the reverse process to the forward edge and not simply some alternative process for moving between the mesostates. The log label ratio, $\ln[\ell(i\ra j)/\ell(j \ra i)]$ is then the total entropy change in taking the transition from $i$ to $j$: the entropy change in the reservoirs together with the internal entropy difference between $j$ and $i$. This form of ``local detailed balance'' goes back to Hill and Schnakenberg and has been broadly justified within stochastic thermodynamics \cite{sei08,bfc15}. Let $R(i,j)$ denote the set of reversible paths, $i = i_1 \lra i_2 \lra \cdots \lra i_k = j$ from $i$ to $j$. If $P \in R(i,j)$ is such a path, let $S(P)$ denote the total entropy change, as above, along the path. Evidently, 
\begin{equation}
S(P) = \ln\left[\left(\frac{\ell (i_1 \to i_2)}{\ell(i_2 \to i_1)}\right) \cdots \left(\frac{\ell (i_{k-1} \to i_k)}{\ell(i_k \to i_{k-1})}\right)\right] \,.
\label{e-spp}
\end{equation}
If $P \in R(i,j)$, let $P^* \in R(j,i)$ denote the reverse path, so that $S(P^*) = -S(P)$.

If the graph can reach thermodynamic equilibrium, an alternative basis element, $\mu(G) \in \ker\lap(G)$, may be found. In this case, detailed balance holds: each pair of reversible edges, $i \lra j$, is in s.s. flux balance, so that $u^*_i(G)\ell(i \ra j) = u^*_j(G)\ell(j \ra i)$. Equivalently, given any cycle of reversible edges, $Q \in R(i,i)$, $S(Q) = 0$. Hence, if $P_1, P_2 \in R(i,j)$, then $S(P_1) = S(P_2)$. We can then define $\mu_i(G) = \exp(-S(P))$ for any $P \in R(i,1)$. As before, $u^*(G) \propto \mu(G)$, which gives the following specification for equilibrium steady-state probabilities,
\begin{equation}
u^*_i(G) \propto \exp(-S(P)) \,,
\label{e-mu}
\end{equation}
for any $P \in R(i,1)$. A similar formula to Eq.\ref{e-ssp} holds, with $\mu$ in place of $\rho$. This formula is the prescription of equilibrium statistical mechanics, with the denominator, $\mu_1(G) + \cdots + \mu_n(G)$, being the partition function for the grand canonical ensemble. 

\subsection*{Reformulating steady-state probabilities}

Path entropies enable the first step in reformulating Eq.\ref{e-mtt}. Following \cite{wcg18}, let $\Phi_{i,j}: \Theta_i(G) \ra \Theta_{j}(G)$ be defined as follows. Choose $T \in \Theta_i(G)$. By construction, there is a unique path in $T$ from $j$ to $i$. Since it has no repeated vertices, this path is minimal. Reversing the edges on this minimal path yields a spanning tree rooted at $j$, which is $\Phi_{i,j}(T) \in \Theta_j(G)$ (Fig.\ref{f-1}C). $\Phi_{i,j}$ is a bijection---there are the same number of spanning trees at each vertex of a reversible graph---and $\Phi_{i,j}^{-1} = \Phi_{j,i}$ \cite{wcg18}. Let $M(i,j) \subseteq R(i,j)$ be the set of minimal paths from $i$ to $j$. While $R(i,j)$ is infinite, $M(i,j)$ is finite. If we focus on the reference vertex and consider any $T \in \Theta_1(G)$, let $T_i \in M(i,1)$ be the unique minimal path, as in the definition of $\Phi_{1,i}$. It is easy to see that \cite{wcg18}, $q(\Phi_{1,i}(T)) = \exp(S(T_i^*))q(T)$. Because $\Phi_{i,j}$ is a bijection, we can rewrite Eq.\ref{e-mtt} for any vertex $i$ in terms of only the spanning trees rooted at $1$. Recalling that $S(T_i^*) = -S(T_i)$, we see that, 
\begin{equation}
\rho_i(G) = \sum_{T \in \Theta_1(G)} \exp(-S(T_i))q(T) \,.
\label{e-mt2}
\end{equation}

We define the arboreal probability distribution on $\Theta_1(G)$ by normalising $q(T)$ to its total over all trees $T$, so that $\pr_{\Theta_1(G)}(T) = q(T)/(\sum_{T \in \Theta_1(G)} q(T))$. This arboreal distribution has been previously studied (Discussion). It follows from Eq.\ref{e-mt2} that $\rho_i(G) \propto \vtx{\exp(-S(T_i)}$, where the average is taken over the arboreal distribution. Since $u^*(G) \propto \rho(G)$ in $\ker\lap(G)$, we see that,
\begin{equation}
u^*_i(G) \propto \vtx{\exp(-S(T_i))}_{\Theta_1(G)} \,.
\label{e-mt3}
\end{equation}
An easy consequence of Eq.\ref{e-mt3} is that
\[ \min(S(T_i)) \leq \ln\left(\frac{u^*_1(G)}{u^*_i(G)}\right) \leq \max(S(T_i)) \,,\]
where the extrema are taken over $T \in \Theta_1(G)$. Maes \emph{et al} derive these bounds by similar means \cite[Cor.2.2]{mkn13} but without the probabilistic rephrasing in Eq.\ref{e-mt3}.

To interpret the arboreal distribution thermodynamically, we express edge labels in Arrhenius form, $\ell(i \ra j) = \exp(\epsilon_i - W_{i \ra j})$. Here, $\epsilon_i$ can be thought of as a vertex energy for mesostate $i$ and $W_{i \ra j}$ as the resulting barrier energy of the edge from $i$ to $j$. In general, $W_{i \ra j} \not= W_{j \ra i}$. Such a representation is always numerically possible but is not unique. Choose any Arrhenius representation and let $T \in \Theta_1(G)$. Let $E(T)$ be the total edge barrier energy, 
\begin{equation}
E(T) = \sum_{i \ra j \in T} W_{i \ra j} \,.
\label{e-et}
\end{equation}
Since $q(T) = \exp(\sum_{1 \leq i \leq n} \epsilon_i)\exp(-E(T))$, and the first term is independent of $T$, the arboreal distribution may be expressed in terms of $E(T)$ as,
\begin{equation}
\pr_{\Theta_1(G)}(T) = \frac{\exp(-E(T))}{\sum_{T \in \Theta_1(G)} \exp(-E(T))} \,.
\label{e-prt}
\end{equation}
Eq.\ref{e-prt} is independent of the choice of Arrhenius rates. It reveals the arboreal distribution to be ``Boltzmann-like'', with the energy of a spanning tree being the total edge barrier energy over the tree. 

Eqs.\ref{e-mt3} and \ref{e-prt} constitute our reformulation of s.s. probabilities. In contrast to the MTT in Eq.\ref{e-mtt}, which lacks thermodynamic meaning, Eq.\ref{e-mt3} smoothly generalises the equilibrium formula in Eq.\ref{e-mu}. At equilibrium, s.s. probabilities are given by path entropies: $u_i^* \propto \exp(-S(P))$. Away from equilibrium, they are given by averages over path entropies: $u_i^* \propto \vtx{\exp(-S(T_i))}$, where the average is calculated over the arboreal distribution. At equilibrium, the entropies of all paths in $R(i,1)$ are identical; the arboreal distribution factors out and Eq.\ref{e-mt3} reduces to Eq.\ref{e-mu}. 

\subsection*{Combinatorial scaling by energetic edges}

Eq.\ref{e-mt3} has a further important advantage. Unlike spanning trees, minimal path entropies do not scale with the size of the graph. Suppose that $G$ satisfies detailed balance and let $\ell_{eq}(i \ra j)$ denote the edge labels under this condition. Suppose that edge labels are altered to break detailed balance and the new labels are given by $\ell(i \ra j) = m(i \ra j)\ell_{eq}(i \ra j)$. We will say that $i \ra j$ is an \emph{energetic edge} if $m(i \ra j) \not= 1$. Let $P:\,v = v_1 \lra \cdots \lra v_k = w \in R(v,w)$ and define $F(P)$ to be the set of energetic edges in the forward direction of $P$,
\begin{equation}
F(P) = \{ v_l \ra v_{l+1} \,|\, m(v_l \ra v_{l+1}) \not= 1 \}\,.
\label{e-fp}
\end{equation}
The set of energetic edges in the reverse direction is then $F(P^*)$. It follows from Eq.\ref{e-spp} that
\begin{equation}
S(P) = \ln\left[\frac{\prod_{i \ra j \in F(P)} m(i \ra j)}{\prod_{i \ra j \in F(P^*)} m(i \ra j)}\right] + S_{eq}(P) \,,
\label{e-llf}
\end{equation}
where $S_{eq}(P)$ is the total entropy change along $P$ at thermodynamic equilibrium. As noted previously, $S_{eq}(P)$ is independent of $P \in R(v,w)$. Given $P_1, P_2 \in R(v,w)$, we will say that $P_1$ is \emph{energetically similar} to $P_2$, denoted $P_1 \sim P_2$, if $F(P_1) = F(P_2)$ and $F(P_1^*) = F(P_2^*)$. It follows from Eq.\ref{e-llf} that if $P_1 \sim P_2$, then $S(P_1) = S(P_2)$. Hence, the number of distinct minimal path entropies in Eq.\ref{e-mt3} is independent of the size of the graph and depends only on the number of energetic edges. This scaling still incurs a combinatorial increase, since minimal paths may have different subsets of energetic edges, but the scaling is substantially less intimidating than that arising from all rooted spanning trees (above). We examine below the implications of this scaling for a graph with a single energetic edge.

\subsection*{Generalised Hopfield discrimination}

\begin{figure}[t]
\begin{center}
\includegraphics[viewport=64 296 524 604,clip=true,width=0.75\textwidth,keepaspectratio]{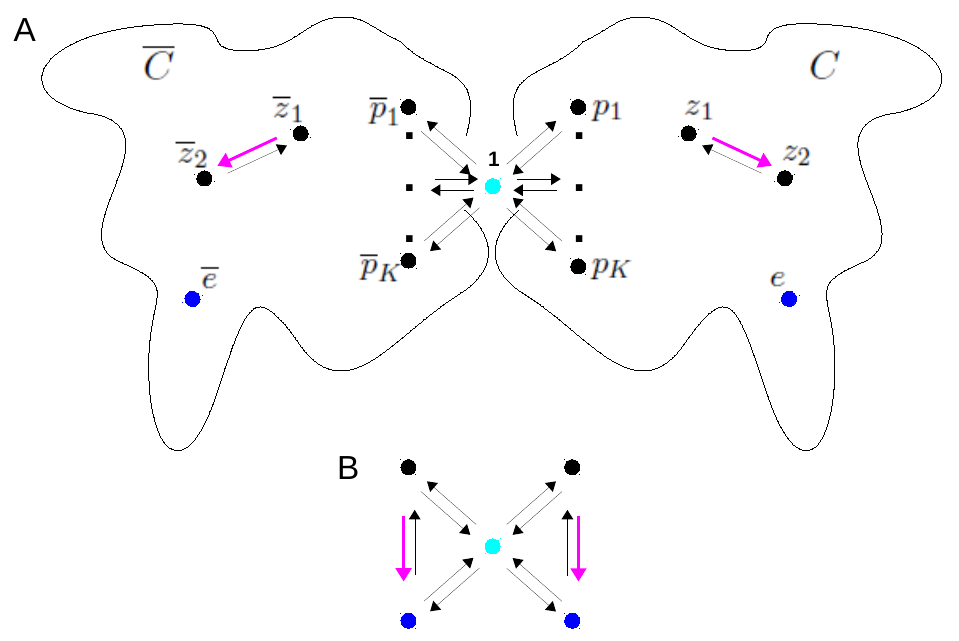}  
\caption{Hopfield discrimination. \textbf{A} Schematic butterfly graph, $G = \ol{C} \bowtie C$, for generalised Hopfield discrimination \cite{wag18}. The subgraphs $C$ and $\ol{C}$ for discriminating the correct and incorrect substrate, respectively, are structurally mirror images with a shared reference vertex (cyan). $C$ is essentially arbitrary (cloud outline---see the text), with $K$ proximal vertices, a single energetic edge (magenta) and an exit vertex (blue). Only the graph structure is shows, with labels omitted. \textbf{B} The butterfly graph structure for Hopfield's original analysis \cite{hop74}.}
\label{f-2}
\end{center}
\end{figure} 

Hopfield's analysis of discrimination between a correct and incorrect substrate sought to explain the low error rates in RNA and DNA synthesis \cite{hop74}. Here, we analyse a general mechanism of Hopfield discrimination using the linear framework approach of \cite{wag18}. Fig.\ref{f-2}A shows a butterfly graph, $G = \ol{C} \bowtie C$, consisting of two ``wings'', $\ol{C}$ and $C$, sharing a common reference vertex, $1$, (cyan). If $C$ and $\ol{C}$ are strongly connected, so too is $G$, and \cite{wag18},
\begin{equation}
\rho_i(G) = \left\{
              \begin{array}{ll}
                \rho_i(\ol{C})\rho_1(C) & \mbox{if $i \in \ol{C}$} \\
                \rho_1(\ol{C})\rho_i(C) & \mbox{if $i \in C$} \,.
              \end{array}\right.
\label{e-btr}
\end{equation}
For Hopfield discrimination, $C$ represents the mesostates interacting with the correct substrate and $\ol{C}$ the same for the incorrect substrate. Structurally (ie: ignoring labels), $C$ and $\ol{C}$ are mirror images of each other. Using overlines to map graph entities in $C$ to their mirror images in $\ol{C}$, $i \ra j$ if, and only if, $\ol{i} \ra \ol{j}$. $C$ is assumed to be reversible and strongly connected but otherwise arbitrary. Ligand binding to vertex $1$ leads to $K$ \emph{proximal vertices}, $p_1, \cdots, p_K$ and ligand is selected to be correct at a distinguished \emph{exit vertex}, $e \not= 1$. The \emph{error ratio} is, 
\begin{equation}
\varepsilon = \frac{u^*_{\ol{e}}(G)}{u^*_e(G)} = \frac{\rho_{\ol{e}}(\ol{C})\rho_1(C)}{\rho_1(\ol{C})\rho_e(C)} \,.
\label{e-vef}
\end{equation}
where the second equality comes from Eqs.\ref{e-ssp} and \ref{e-btr}. Assume to begin with that $G$ is at thermodynamic equilibrium with labels $\ell_{eq}(i \ra j)$. Following Hopfield, discrimination only takes place through unbinding from proximal vertices. Accordingly, $\ell_{eq}(i \ra j) = \ell_{eq}(\ol{i} \ra \ol{j})$ as long as $j \not= 1$ and $\ell_{eq}(\ol{p}_u \ra 1) = \alpha\ell_{eq}(p_u \ra 1)$, where $\alpha > 1$, so that the incorrect substrate has a higher off rate. Using Eq.\ref{e-mu}, it follows that the equilibrium error ratio is $\varepsilon_{eq} = \alpha^{-1}$. Accordingly, $\eps_{eq} < 1$.

Hopfield's insight was that $\varepsilon_{eq}$ is independent of the number of discriminations, $K$, and the only way to exceed this ``Hopfield barrier'' \cite{edg16} is to expend energy. He analysed the graph in Fig.\ref{f-2}B, for which $K = 2$ and $C$ has only three vertices, with energy expenditure on the magenta edge, and identified a parametric region for kinetic proofreading in which $(\eps_{eq})^2 < \eps < \eps_{eq}$ \cite{hop74,wag18}. The question we ask is what determines $\varepsilon$ when $K > 2$ and $C$ is a general graph (Fig.\ref{f-2}A) in which energy is expended at only a single energetic edge (magenta) where $\ell(z_1 \ra z_2) = m\ell_{eq}(z_1 \ra z_2)$.

We calculate how $\varepsilon$ depends on $m$ and $\alpha$ by exploiting the reformulation above, with Eq.\ref{e-mt2} being more convenient for this purpose than Eq.\ref{e-mt3}, and by partitioning trees according to proximal edges (ie: edges $j \ra 1$, where $j$ is a proximal vertex). We give a sketch here, with details in the Materials and Methods (M \& M).

Given a polynomial ${\cal P}$ in the edge labels, we say that it is $m$-free, respectively $\alpha$-free, if $m$, respectively $\alpha$, does not divide any monomial in ${\cal P}$. Let $\Theta = \Theta_1(C)$. Eq.\ref{e-mt2} leads us to partition $\Theta$ according to the combinatorics of the energetic edge on minimal paths in $M(e,1)$, 
\begin{equation}
\begin{array}{lcl}
\Theta_0 & = & \{T \in \Theta \,|\, z_1 \ra z_2, z_2 \ra z_1 \not\in T_e \} \\
\Theta_+ & = & \{T \in \Theta \,|\, z_2 \ra z_1 \in T_e \}\\
\Theta_- & = & \{T \in \Theta \,|\, z_1 \ra z_2 \in T_e \} \,.
\end{array}
\label{e-th0}
\end{equation}
Let $P_0, P_+, P_- \in M(e,1)$ be any choices of minimal paths arising as $T_e$ for $T \in \Theta_0, \Theta_+, \Theta_-$, respectively. By construction, different choices are energetically similar, so that $S(P_0), S(P_+), S(P_-)$ are well defined for any choices of minimal paths. Furthermore, by Eq.\ref{e-llf}, $S(P_+) = S(P_0) - \ln(m)$, $S(P_-) = S(P_0) + \ln(m)$. Let us extend $q$ to subsets $X \subseteq \Theta$ by defining $q(X) = \sum_{T \in X}q(T)$ and note that $q_{eq}$ corresponds to $m = 1$. We see from Eq.\ref{e-th0} that $q(\Theta_+)$ is $m$-free and that $q(\Theta_-) = m\,q_{eq}(\Theta_-)$. Hence we can rewrite Eq.\ref{e-mt2} as,
\begin{equation}
\rho_e(C) = \exp(-S(P_0))\left(q(\Theta_0) + m\,q(\Theta_+) + q_{eq}(\Theta_-)\right) \,,
\label{e-rec}
\end{equation}
where $q(\Theta_0)$ and $q_{eq}(\Theta_-)$ are $m$-free but $q(\Theta_0)$ may not be. The three parts of Eq.\ref{e-rec}, which come from the tripartite combinatorics of Eq.\ref{e-th0}, reflect the presence of only a single energetic edge.

We now introduce the partitioning scheme. Given $X \subseteq \Theta$, let $X^{(j,u)} \subseteq X$, for $1 \leq j \leq K$ and $0 \leq u \leq 1$, consist of those trees in $X$ with exactly $j$ proximal edges to $1$ and exactly $u$ energetic edges. The $X^{(j,u)}$ form a partition of $X$ into mutually disjoint subsets, so that $q(X) = \sum_{j,u} q(X^{(j,u)})$. By construction, $q(X^{(j,0)})$ is $m$-free and $q(X^{(j,1)}) = m\,q_{eq}(X^{(j,1)})$, where $q_{eq}(X^{(j,1)})$ is $m$-free. Most importantly, again by construction,
\begin{equation}
q(\ol{X}^{(j,u)}) = \alpha^j q(X^{(j,u)}) \,,
\label{e-qox}
\end{equation}
where, evidently, $q(X^{(j,u)})$ is $\alpha$-free. 

Eqs.\ref{e-rec} and \ref{e-qox} make it straightforward to calculate the error ratio from Eq.\ref{e-vef}, 
\begin{equation}
\varepsilon = \varepsilon_{eq}\left(\frac{(\ol{P}m + \ol{Q})(Rm + S)}{(\ol{R}m + \ol{S})(Pm + Q)}\right) \,,
\label{e-epq}
\end{equation}
where the eight coefficients in Eq.\ref{e-epq}, which are all $m$-free, are expressed in terms of the constructions above in Eq.\ref{e-25}. It is striking that Eq.\ref{e-epq} has the same algebraic form for the general graph in Fig.\ref{f-1}A as for Hopfield's simple graph in Fig.\ref{f-1}B \cite[Eq.4]{wag18}, albeit with vastly more complicated coefficients. The overlined coefficients in Eq.\ref{e-epq} are each of degree $K$ in $\alpha = \varepsilon_{eq}^{-1}$, so that $\varepsilon$ is a rational function of $\varepsilon_{eq}$ and $m$. Eq.\ref{e-epq} enables us to analyse discrimination in complex graphs (below).

Murugan \emph{et al}, using Schnakenberg's version of Eq.\ref{e-mtt}, showed the rational dependence of $\varepsilon$ on $\varepsilon_{eq}$ \cite[Eq.10]{mhl14}. Their treatment was based on a network similar to the butterfly graph in Fig.\ref{f-2}A, with certain structural restrictions---the number of edges leaving the reference vertex equals the number entering the exit vertex---but allows for discriminations at non-proximal edges and global energy expenditure \cite[Fig.3]{mhl14}. In view of the latter, they were unable to find an expression for $\varepsilon$, as we have in Eq.\ref{e-epq}. They did observe the following bounds on $\varepsilon$, 
\begin{equation}
(\eps_{eq})^K < \eps < (\eps_{eq})^{2-K} \,,
\label{e-vek}
\end{equation}
where the quantity corresponding to our $K$ is the number of ``discriminatory edges'' \cite[C(i)]{mhl14}. Eq.\ref{e-vek} is easy to deduce from Eq.\ref{e-epq}. The left-hand inequality in Eq.\ref{e-vek} generalises Hopfield's finding for Fig.\ref{f-1}B with $K = 2$. As noted by Murugan \emph{et al}, Eq.\ref{e-vek} allows for ``anti-proofreading'' regimes, where energy expenditure worsens the error ratio above the equilibrium value \cite{mhl14}. Indeed, this is already seen in the quadratic dependency of $\varepsilon$ on $m$ in Eq.\ref{e-epq} \cite{wag18}. 


\subsection*{Other previous work and optimal discrimination}

There have been other studies of discrimination in addition to \cite{mhl14}. We have already drawn on \cite{wag18} for the butterfly graph in Fig.\ref{f-2}A. This work developed an asymptotic approach to the tradeoff between accuracy and speed of discrimination. Ehrenberg and colleagues \cite{ebl80,beh81} and Savageau and colleagues \cite{fsa80,sla81} rigorously analysed Hopfield's remarks on multi-stage proofreading schemes \cite{hop74}. In our language, they vertically extended the graph in Fig.\ref{f-2}B to have multiple triangular ``stages''. These studies established conditions for minimising energy expenditure for a given error ratio. Murugan \emph{et al}, along with the results noted above, also analysed the accuracy-speed tradeoff \cite{mhl12,mhl14}, finding multiple proofreading regimes in general networks similar to the butterfly graph in Fig.\ref{f-2}A. They stated that optimal discrimination, for which $(\eps_{eq})^K < \eps < (\eps_{eq})^{K-1}$, can be achieved in such networks for appropriate parameter choices \cite[Eq.3]{mhl12}. An important distinction between these studies is that Murugan \emph{et al} allow energy expenditure anywhere in the network, while Ehrenberg \emph{et al} and Savageau \emph{et al} allow energy expenditure at only one transition, as we have done above. We were therefore interested to know whether optimal discrimination is still possible in complex graphs with only a single energetic edge.

\begin{figure}[t]
\begin{center}
\includegraphics[viewport=76 180 536 378,clip=true,width=0.75\textwidth,keepaspectratio]{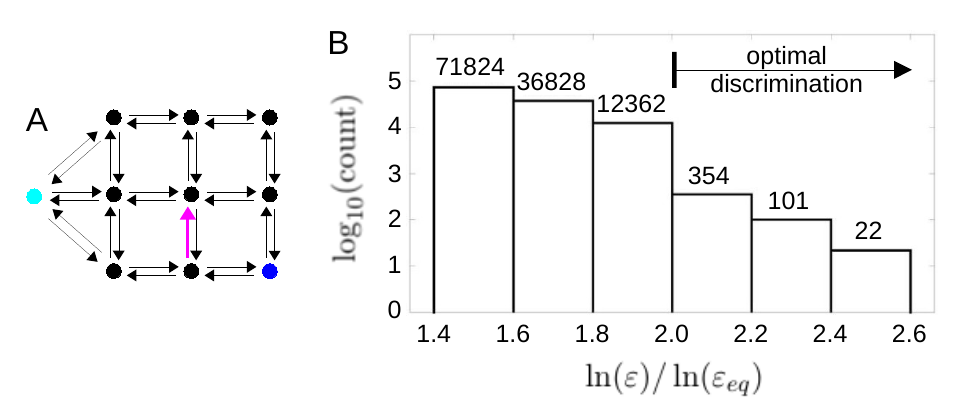}  
\caption{Optimal Hopfield discrimination in a complex graph. \textbf{A} Right-hand (correct) wing of butterfly graph structure, with $K = 3$ proximal vertices, following the same conventions as Fig.\ref{f-2}, with a single energetic edge (magenta) separated from the proximal vertices and the exit vertex (blue). \textbf{B} Histogram of numerical calculations for the graph in \textbf{A} giving logarithmic counts of randomly sampled parameter sets, with actual numbers over each bar, for specified ranges of normalised logarithmic error ratios. Those parameter sets achieving optimal discrimination, with $(\eps_{eq})^3 < \eps < (\eps_{eq})^2$, are indicated. The calculations used Eq.\ref{e-epq} (Materials and Methods).}
\label{f-3}
\end{center}
\end{figure} 

We found that the answer is yes. For example, consider the graph in Fig.\ref{f-3}A. We note that this graph does not meet the structural restrictions mentioned above for the networks studied by Murugan \emph{et al} in \cite{mhl12,mhl14}. We exploited the error ratio formula in Eq.\ref{e-epq} to identify sets of parameter values for which this graph exhibits optimal discrimination, as detailed in Fig.\ref{f-3}B (M \& M). Such discrimination requires that each proximal edge contributes in a synergistic manner to the reduction in error, despite energy being expended at only a single edge which is distant in the graph. This synergy presumably arises from the global dependence of steady-state probabilities on all edge labels. It suggests remarkable global functional capabilities, despite only local energy expenditure, when the right parametric conditions are satisfied. It remains an intriguing problem to determine when optimal discrimination occurs and we hope to report on this subsequently.

\subsection*{The entropy production index}

We have studied Hopfield discrimination under the assumption of a single energetic edge. This relies, however, on the prior choice of a state of thermodynamic equilibrium. Moreover, energy expenditure could also occur not simply at one edge but through coupling at multiple edges to reservoirs with different chemical potentials, as happens during enzymatic catalysis. It is important, therefore, to have a more independent and flexible means to exploit the reformulation in Eq.\ref{e-mt3}. In the spirit of Schnakenberg, \cite{sch76}, we consider any basis, $B$, of oriented minimal cycles, $B = \{B_1, \cdots, B_q\}$. Such a basis arises by choosing a spanning tree and adjoining an edge outside the tree to create each minimal cycle, for which an arbitrary orientation is chosen (M \& M). The number, $q$, of minimal cycles in any basis is the Betti number of $G$, $\beta_1(G)$, or the rank of the first homology group of $G$ considered as a topological space. Let $\iota(B)$ denote the number of cycles $B_i$ which break the cycle condition, $\iota(B) = \# i\,, 1 \leq i \leq q\,, S(B_i) \not= 0$. We define the entropy production index of the graph $G$, $\iota(G)$, to be the minimum of these quantities over all cycle bases, $\iota(G) = \min_{B} \iota(B)$. Evidently, $0 \leq \iota(G) \leq \beta_1(G)$. The case $\iota(G) = 0$ corresponds to thermodynamic equilibrium. It is not difficult to show that $\iota(G) = 1$ corresponds to the case considered here, and also in \cite{ebl80,beh81,fsa80,sla81}, of a single energetic edge (M \& M). The case considered in \cite{mhl12,mhl14} corresponds to the entropy production index being maximal, $\iota(G) = \beta_1(G)$. The quantity $\iota(G)$, which is thermodynamic in the language of Maes \cite{mae20} (below), offers a first step in partitioning the non-equilibrium landscape. 

\section*{DISCUSSION}

The Matrix-Tree theorem (MTT) gives an exact solution, in terms of the transition rates, for the s.s. probabilities of a Markov process (Eq.\ref{e-mtt}). The MTT's elegant mathematical statement belies its intractability. Away from thermodynamic equilibrium, the s.s. probability of a vertex, $i$, is globally dependent on all edge labels in the graph, in stark contrast to equilibrium, in which only the edge labels on a minimal path in $M(1,i)$ are relevant (Eq.\ref{e-mu}). At equilibrium, s.s. probabilities are path independent; away from equilibrium they are not merely path dependent but every path in $M(1,i)$ is needed and the MTT does the bookkeeping for this calculation by way of spanning trees (Eq.\ref{e-mtt}). This requires enumerating all spanning trees rooted at $i$, leading to a combinatorial explosion that leaves even simple graphs beyond the reach of calculation. Studies have had to rely, in effect, on astute approximations to a few dominant trees. While this has provided important insights it also suggests that those behaviours which depend on small contributions from many trees may have been overlooked. This could be a particularly serious omission in biology, where functionality can arise from many small contributions \cite{cms15}.

We have overcome the intractability of s.s. probabilities in two ways. First, by recasting the MTT in thermodynamic terms as a generalisation of the equilibrium case: in place of minimal path entropies at equilibrium (Eq.\ref{e-mu}), averages of these quantities must be taken away from equilibrium (Eq.\ref{e-mt3}). Here, the average is calculated over the Boltzmann-like probability distribution on spanning trees which we call the arboreal distribution (Eq.\ref{e-prt}). Second, energetically similar minimal paths have identical entropies so that the number of distinct minimal path entropies scales with the number of energetic edges, not the size of the graph. It is this scaling which has enabled exact calculation of the error ratio (Eq.\ref{e-epq}), for an arbitrary graph with only a single energetic edge (Fig.\ref{f-2}A). Such a calculation would not have been feasible with the un-reformulated MTT \cite{mhl14}.

Expressions for steady-state probabilities which bear resemblance to Eq.\ref{e-mt3} have been previously described, as in \cite[Eq.15]{knn08} and \cite[Eq.3.12]{mkw08}. These lack, however, the arboreal distribution on spanning trees. This distribution is important, in our view, because it gives an exact description and is Boltzmann-like (Eq.\ref{e-prt}). Furthermore, it has previously been studied mathematically, although not given a name, as the fixed point of algorithms for generating random spanning trees. This goes back to work of Broder and Aldous \cite{abr89,dal90}, which was extended to the labelled (``weighted'') graphs used here by Wilson \cite{wil96}; see \cite{pwt18} for a recent discussion. Perhaps most importantly, the arboreal distribution cleanly separates local and global contributions. Landauer pointed out the necessity for kinetic, non-thermodynamic quantities to exactly describe the global nature of non-equilibrium steady states \cite{rla75}, despite significant attempts to characterise them thermodynamically \cite{etj57,ipr67}. Landauer's point has been reiterated by Maes in his discussion of ``frenetic'' behaviour \cite{mae20}. Eq.\ref{e-mt3} separates the local, thermodynamic contribution to steady-state probabilities, $\exp(-S(T))$, from the kinetic, global contribution coming from the arboreal distribution, $\pr_{\Theta_1(G)}$. Our reformulation offers, therefore, a principled and exact description of non-equilibrium steady states within the enclosed garden of Markov processes. An important task for subsequent work is to characterise the arboreal distribution, for which the mathematical connections mentioned above may be helpful. 

The reformulation in Eq.\ref{e-mt3} is also appealing because of its formal resemblance to other path ensemble formulations in physics. In quantum mechanics, the probability amplitude for a particle going from $A$ to $B$ is the integral over all paths from $A$ to $B$ of the action along the paths \cite{fey48}. In statistical mechanics, equilibrium information can be recovered from driven non-equilibrium paths by averaging the exponential of the work performed along the paths \cite{jar97b} and many exact non-equilibrium fluctuation theorems can be obtained in this way \cite{cro00}. Interestingly, in the present paper, it is non-equilibrium information which is recovered from a Boltzmann-like distribution. Whether some more fundamental setting underlies these different path ensemble formulations is beyond the scope of this paper but we note how essential the graph-theoretic framework is to clarifying the ensemble. We cannot speak of ``paths'' without the graph and it is the graph which yields the spanning trees on which the arboreal distribution is defined. The reformulation reinforces the central role of the graph in the non-equilibrium behaviour of Markov processes, testifying again to the pioneering insights of Hill and Schnakenberg. 

The graph further clarifies Hopfield's analysis of kinetic proofreading \cite{hop74}. At equilibrium, path independence (Eq.\ref{e-mu}) implies that the system cannot tell how many discriminations have been made, so that $\eps_{eq}$ is independent of their number. Away from equilibrium, global path dependency permits, in principle,  profoundly different behaviour, in which energy dissipation, even locally at a single edge, can enable multiple discriminations to collectively reduce the error (Fig.\ref{f-3}).

Finally, an important message of the present paper is that systems for which $\iota(G) = 1$, while suffering all the global parameter dependency and combinatorial complexity that arise away from thermodynamic equilibrium, are nevertheless substantially simpler to analyse than those with higher entropy production index, as Eq.\ref{e-epq} and Fig.\ref{f-3} testify. It is $\iota(G)$, not the size of the graph, which determines the number of distinct minimal path entropies that must be dealt with in Eq.\ref{e-mt3}. When $\iota(G) = 1$, there are only 3 such entropies, which leads to the compact and tractable form of Eq.\ref{e-rec}. Being ``away from thermodynamic equilibrium'' is not an unitary condition but, rather, a nuanced and complex landscape, in which how and where energy is expended can profoundly influence functional outcomes. The ideas introduced here suggest how we can begin to ``follow the energy'' through this non-equilibrium landscape to uncover the logic of biological information processing.

\section*{Materials and Methods}
\subsection*{Partitioning scheme}

Recall the partitioning scheme introduced above in which a subset of spanning trees rooted at the reference vertex, $X \subseteq \Theta$, is divided into parts, $X^{(j,u)}$, for $1 \leq j \leq K$ and $0 \leq u \leq 1$. $X^{(j,u)}$ consists of those trees in $X$ with exactly $j$ proximal edges to $1$ and exactly $u$ energetic edges. In other words, $T \in X^{(j,u)}$ if, and only if, there are distinct proximal vertices, $p_{i_1}, \cdots, p_{i_j}$ in $C$, which may depend on $T$, such that $p_{i_1} \ra 1, \cdots, p_{i_j} \ra 1 \in T$ but no other proximal vertices have this property; and, if $u = 0$, $z_1 \ra z_2 \not\in T$, while if $u = 1$, $z_1 \ra z_2 \in T$. Note that any tree in $\Theta$ must have between $1$ and $K$ proximal edges to $1$ and no more than $1$ energetic edge (Fig.\ref{f-2}A). It follows that the $X^{(j,u)}$ form a partition of $X$ into disjoint subsets, 
\begin{equation}
X = (X^{(1,0)} \amalg X^{(1,1)}) \amalg \cdots \amalg (X^{(K,0)} \amalg X^{(K,1)}) \,.
\label{e-11}
\end{equation}
Hence, we can decompose $q(X)$ as follows, 
\begin{equation}
\begin{split}
q(X) & =  \left(q(X^{(1,0)}) + \cdots + q(X^{(K,0)})\right) \\
& + m\left(q_{eq}(X^{(1,1)}) + \cdots + q_{eq}(X^{(K,1)})\right)
\end{split}
\label{e-18}
\end{equation}
where the terms $q(X^{(j,0)})$ and $q_{eq}(X^{(j,1)})$ are all $m$-free. The value of this decomposition becomes clear by applying Eq.\ref{e-qox} to see that,
\begin{equation}
\begin{split}
q(\ol{X}) & = \left(\alpha q(X^{(1,0)}) + \cdots + \alpha^K q(X^{(K,0)})\right) \\
& + m\left(\alpha q_{eq}(X^{(1,1)}) + \cdots + \alpha^K q_{eq}(X^{(K,1)})\right) \,,
\end{split}
\label{e-19}
\end{equation}
where the expressions $q(X^{(j,0)})$ and $q_{eq}(X^{(j,1)})$ are both $m$-free, as above, and, by construction, $\alpha$-free. 

\subsection*{Calculation of $\rho_1(C)$ and $\rho_1(\ol{C})$}

Let us abbreviate $\sum_{1 \leq j \leq K}$ by $\sum_j$. Since $\rho_1(C) = q(\Theta)$, it follows from Eq.\ref{e-18} that,
\begin{equation}
\rho_1(C) = \left(\sum_j q(\Theta^{(j,0)})\right) + m \left(\sum_j q_{eq}(\Theta^{(j,1)}) \right) \,.
\label{e-22}
\end{equation}
Similarly, applying Eq.\ref{e-19} to $\rho_1(\ol{C}) = q(\ol{\Theta})$, we see that,
\begin{equation}
\rho_1(\ol{C}) = \left(\sum_j \alpha^j q(\Theta^{(j,0)})\right) + m \left(\sum_j \alpha^j q_{eq}(\Theta^{(j,1)}) \right) \,.
\label{e-23}
\end{equation}

\subsection*{Calculation of $\rho_e(C)$ and $\rho_{\ol{e}}(\ol{C})$}

We begin with Eq.\ref{e-rec} which represented $\rho_e(C)$ in terms of a partition into disjoint subsets of $\Theta$, 
\begin{equation}
\Theta = \Theta_0 \amalg \Theta_+ \amalg \Theta_- \\
\label{e-16}
\end{equation}
It is clear from Eq.\ref{e-rec} that, for all $1 \leq j \leq K$, 
\begin{equation}
\Theta_+^{(j,1)} = \emptyset \hspace{0.5em}\mbox{and}\hspace{0.5em} \Theta_-^{(j,0)} = \emptyset \,.
\label{e-27}
\end{equation}
If we now apply the decomposition in Eq.\ref{e-18} to each of the three subsets of Eq.\ref{e-16}, and collect together the $m$-free terms in Eq.\ref{e-rec}, we find that, 
\begin{equation}
\begin{array}{c}
\rho_e(C) = \exp(-S(P_0)) \,\,\times \\
\left[\sum_j \left(q(\Theta_0^{(j,0)}) + q_{eq}(\Theta_-^{(j,1)}) \right)\right. \\
+ \left. m \left( \sum_j \left(q_{eq}(\Theta_0^{(j,1)}) + q(\Theta_+^{(j,0)})\right)\right)\right] \,.
\end{array}
\label{e-17}
\end{equation}
Similarly, using the decomposition in Eq.\ref{e-19}, we find that,
\begin{equation}
\begin{array}{c}
\rho_{\ol{e}}(\ol{C}) = \alpha^{-1}\exp(-S(P_0)) \,\,\times \\
\left[ \sum_j \alpha^j\left(q(\Theta_0^{(j,0)}) + q_{eq}(\Theta_-^{(j,1)}) \right) \right. \\
+ \left. m \left( \sum_j \alpha^j\left(q_{eq}(\Theta_0^{(j,1)}) + q(\Theta_+^{(j,0)})\right)\right) \right] \,,
\end{array}
\label{e-21}
\end{equation}
where we have used the fact that $S(\ol{P_0}) = \alpha^{-1}S(P_0)$. 

\subsection*{Proof of Eq.\ref{e-epq}}

We have now calculated each of the four terms appearing in Eq.\ref{e-epq}. Using overlines again to indicate coefficients from the mirror image subgraph $\ol{C}$ and recalling that $\alpha = (\eps_{eq})^{-1}$, we find that, 
\begin{equation}
\eps = \eps_{eq}\left(\frac{(\ol{P}m + \ol{Q})(Rm + S)}{(\ol{R}m + \ol{S})(Pm + Q)}\right) \,.
\label{e-24}
\end{equation}
The eight coefficients in Eq.\ref{e-24} are given by
\begin{equation}
\begin{array}{lcl}
P & = & \sum_j q_{eq}(\Theta_0^{(j,1)}) + q(\Theta_+^{(j,0)}) \\
Q & = & \sum_j q(\Theta_0^{(j,0)}) + q_{eq}(\Theta_-^{(j,1)}) \\
R & = & \sum_j q_{eq}(\Theta^{(j,1)}) \\
S & = & \sum_j q(\Theta^{(j,0)}) \\
\ol{P} & = & \sum_j \alpha^j (q_{eq}(\Theta_0^{(j,1)}) + q(\Theta_+^{(j,0)})) \\
\ol{Q} & = & \sum_j \alpha^j (q(\Theta_0^{(j,0)}) + q_{eq}(\Theta_-^{(j,1)})) \\
\ol{R} & = & \sum_j \alpha^j q_{eq}(\Theta^{(j,1)}) \\
\ol{S} & = & \sum_j \alpha^j q(\Theta^{(j,0)}) \,.
\end{array}
\label{e-25}
\end{equation}
The various expressions of the form $q(-)$ or $q_{eq}(-)$ in Eq.\ref{e-25} are all both $\alpha$-free and $m$-free. Eqs.\ref{e-24} and \ref{e-25} establish Eq.\ref{e-epq}.

\subsection*{Proof of Eq.\ref{e-vek}.} Using the fact that $\alpha > 1$, which encodes the distinction between the correct and incorrect substrate, we see by inspection of Eq.\ref{e-25}, that
\[
\begin{array}{rcccl} 
\alpha(P + Q) & < & \ol{P} + \ol{Q} & < & \alpha^K(P + Q) \\
\alpha(R + S) & < & \ol{R} + \ol{S} & < & \alpha^K(R + S) \,.
\end{array}
\]
It follows that
\begin{equation}
\alpha < \frac{\ol{P} + \ol{Q}}{P + Q} < \alpha^K
\label{e-pq}
\end{equation}
and also that
\begin{equation}
\alpha^{-K} < \frac{R + S}{\ol{R} + \ol{S}} < \alpha^{-1}
\label{e-rs}
\end{equation}
Combining Eqs.\ref{e-pq} and \ref{e-rs} and using Eq.\ref{e-24} and the fact that $\alpha = (\eps_{eq})^{-1}$, we see that, 
\begin{equation}
(\eps_{eq})^K < \eps < (\eps_{eq})^{2-K} \,,
\label{e-eps}
\end{equation}
which proves Eq.\ref{e-vek}.

\subsection*{Fig.\ref{f-3}B} Eq.\ref{e-epq} expresses $\eps$ as a quadratic rational function of $m$. This graph of this function can exhibit many shapes \cite{wag18} and we looked for the conditions under which it has a positive minimum, which implies proofreading. The derivative, $d\eps/dm$, has a quadratic numerator and it is readily seen that $\eps$ is decreasing at $m = 0$ if,
\begin{equation}
  \frac{S\ol{P}+R\ol{Q}}{P\ol{S}+Q\ol{R}} < \frac{S\ol{Q}}{Q\ol{S}} \,.
\label{e-sop}
\end{equation}
When Eq.\ref{e-sop} holds, $\eps$ has a single positive minimum if also, 
\begin{equation}
  \frac{S\ol{P}+R\ol{Q}}{P\ol{S}+Q\ol{R}} <  \frac{R\ol{P}}{P\ol{R}} \,.
\label{e-rop}
\end{equation}
We sought this minimum error ratio for the graph in Fig.\ref{f-3}A. First, an independent set of edge labels were chosen using a spanning tree, as previously described \cite{edg16}. Numerical values for these rates on the correct wing of the graph, $C$, were chosen independently as $10^x$, where $x$ was drawn randomly from the uniform distribution on $[-3,+3]$. The remaining edge labels, not on the spanning tree, which determine independent cycles, were chosen to make $\iota(G) = 0$, so that $G$ is at thermodynamic equilibrium. The labels on the incorrect wing, $\ol{C}$, were then determined by the relationships described in the main text. The value of $\alpha$, which gives the equilibrium error ratio, $\eps_{eq} = \alpha^{-1}$, was arbitrarily set to $0.1$. Departure from thermodynamic equilibrium was imposed through the multiplier, $m$, on the energetic edge, $\ell(z_1 \ra z_2) = m\ell_{eq}(z_1 \ra z_2)$, while all other edge labels kept their previously assigned values, $\ell(i \ra j) = \ell_{eq}(i \ra j)$ when $i \not= z_1$ or $j \not= z_2$. 
Spanning trees were enumerated using Matlab's \texttt{generateSpanningTrees} function and the eight coefficients, $P, Q, R, S, \ol{P}, \ol{Q}, \ol{R}, \ol{S}$ in Eq.\ref{e-epq} were numerically calculated from Eq.\ref{e-25}. If they did not satisfy the inequalities in Eqs.\ref{e-sop} and \ref{e-rop}, the parameter set was rejected. If they did, the value of $m$ giving the positive minimum of $\eps$ was calculated from the quadratic numerator of $d\eps/dm$ as,
\[ m_* = \frac{P\ol{Q}S\ol{R}-Q\ol{P}R\ol{S} + \sqrt{(P\ol{Q}-Q\ol{P})(SP-QR)(\ol{S}\ol{P}-\ol{Q}\ol{R})(R\ol{S}-S\ol{R})}}{R\ol{P}(P\ol{S}+Q\ol{R}) - P\ol{R}(S\ol{P}+R\ol{Q})} \,,\]
and the corresponding minimum value of $\eps$ was determined by substituting $m_*$ into Eq.\ref{e-epq}. $\sim\!\!10^7$ parameter sets were sampled, of which $\sim\!\!1.2 \times 10^5$ had minimum error ratios satisfying $\ln(\eps)/\ln(\eps_{eq}) > 1.4$, as reported in Fig.\ref{f-3}B. 

\subsection*{Cycle basis and $\mathbf{\iota(G) = 1}$.} Assuming that $G$ is reversible, it is simpler to work with the corresponding undirected and unlabelled graph, $G^u$, in which $i \sim j$ in $G^u$ if, and only if, $i \lra j$ in $G$. An undirected spanning tree $T$ in $G^u$ is a connected, acyclic subgraph that includes each vertex. Choosing any edge which is not in $T$ defines a minimal cycle in $G^u$. Such a cycle can be lifted to a minimal reversible cycle in $G$ and an arbitrary orientation around the cycle may be chosen. Doing this for each minimal cycle obtained from $T$ defines a basis of oriented minimal cycles. If $G$ has $v$ vertices and $e$ reversible pairs of edges, then $T$ has $v-1$ edges and the number of minimal cycles in a basis is the first Betti number of $G$, $\beta_1(G) = e - v + 1$. If $\iota(G) = 1$, then there is a basis of oriented cycles in which only one cycle breaks the cycle condition. Take the undirected edge on the cycle which is not on the corresponding spanning tree in $G^u$. By changing the label of either of the corresponding reversible edges in $G$, the oriented cycle can always be returned to equilibrium. Conversely, suppose $G$ is moved from thermodynamic equilibrium by changing just $\ell(i \ra j)$. Let $u$ denote the undirected edge $i \sim j$ in $G^u$. Take any basis, $B$, of oriented minimal cycles in $G$ and let $T$ be the corresponding undirected spanning tree in $G^u$. If $\iota(B) > 1$, so that multiple cycles in $B$ break the cycle condition, then $u$ must be an edge in $T$, for otherwise it would occur on only one cycle. Choose any cycle on which $u$ appears and suppose that $v$ is the corresponding edge which is not in $T$. Removing $u$ from $T$ and adjoining $v$ creates a new spanning tree, $T'$, in $G^u$. Because $u$ is not in $T'$ by construction, the basis, $B'$, of minimal cycles corresponding to $T'$ has only the cycle defined by $u$ away from equilibrium. Hence, $\iota(B') = 1$ and so $\iota(G) = 1$. It follows that having a single edge away from equilibrium corresponds exactly to $\iota(G) = 1$.

\section*{ACKNOWLEDGEMENTS}
We thank Christian Maes and Kee-Myoung Nam for helpful comments. U.C. was supported by the Giovanni Armenise Harvard Foundation. U.C. and J.G. were also supported by US NSF award \#1462629 and by US NIH award \#R01GM105375.

\bibliography{/home/jhcg/work/BIB/bio}

\end{document}